\documentclass[pre,aps,twocolumn,floatfix,showpacs]{revtex4-1}
\pdfoutput=1
\usepackage{microtype}
\usepackage[british]{babel}

\usepackage{amsmath}
\usepackage{amssymb}
\usepackage{amsfonts}

\usepackage{multirow} 
\usepackage{array}

\usepackage{graphicx}
\usepackage{tikz}
\usetikzlibrary{arrows}

\usepackage[linktocpage]{hyperref}

\newcommand{\thsep}{\thinspace}
\newcommand{\zp}{\;\tikz[baseline=(C.base)]\node[draw,circle,inner sep=0.5pt](C){\small{z}};\;}
\newcommand{\ket}[1]{|{#1}\rangle}
\newcommand{\bra}[1]{\langle{#1}|}

\newcolumntype{C}[1]{>{\centering\arraybackslash}p{#1}}

\graphicspath{{.}}

\begin{document}

\title[Divide \& conquer translations]{Divide and conquer the Hilbert
  space of translation-symmetric spin systems}

\author{Alexander Wei{\ss}e} 
\affiliation{Max-Planck-Institut f{\"u}r Mathematik, 
  Vivatsgasse 7, 53111 Bonn, Germany}
\email{weisse@mpim-bonn.mpg.de}

\date{October 5, 2012, revised: January 14, 2013}

\begin{abstract}
  Iterative methods that operate with the full Hamiltonian matrix in
  the untrimmed Hilbert space of a finite system continue to be
  important tools for the study of one- and two-dimensional quantum
  spin models, in particular in the presence of frustration. To reach
  sensible system sizes such numerical calculations heavily depend on
  the use of symmetries. We describe a divide-and-conquer strategy for
  implementing translation symmetries of finite spin clusters, which
  efficiently uses and extends the ``sublattice coding'' of
  H.\,Q.~Lin. With our method, the Hamiltonian matrix can be
  generated on-the-fly in each matrix vector multiplication, and
  problem dimensions beyond $10^{11}$ become accessible.
\end{abstract}
 
% 02.70.-c  Computational techniques; simulations
% 02.20.-a  Group theory
% 75.10.Jm  Quantized spin models, including quantum spin frustration 
% 05.30.-d  Quantum statistical mechanics


\maketitle

\section{Introduction}
Lattice spin models have attracted continuous research activity, from
the early days of quantum mechanics until the present. Unfortunately,
only very few spin models can be solved analytically and in the
thermodynamic limit, with geometries usually restricted to one
dimension~\cite{Bax82,MC10}. Numeric methods, therefore, are
indispensable for an understanding of quantum spin models. Density
matrix renormalisation~\cite{Wh92} and related variational approaches
have revolutionised the study of one-dimensional systems and are able to
deliver very precise results for eigenstates and correlation
functions. However, two-dimensional spin systems, systems with
frustrated interactions~\cite{LMM11}, and dynamic correlations in such
systems are still a domain for ``exact'' iterative methods that
operate with the full Hamiltonian matrix of a finite cluster. Taking
advantage of symmetries increases the accessible cluster size,
which is crucial for significant results.

In this work we consider spin models on finite lattices with periodic
boundary conditions and describe an efficient approach for the
construction of a translation symmetric basis of the corresponding
Hilbert space.  The underlying ideas are related to divide-and-conquer
strategies and fast Fourier transform. The core decomposition trick we
use was invented more than two decades ago by H.\,Q.~Lin~\cite{Li90},
but for unknown reasons did not really catch on. For many years the
symmetrised Hilbert space dimensions reached in studies of quantum
spin models therefore lagged behind, compared to simulations of
Hubbard-type models or electron-phonon models.

\section{Standard approach}\label{sec:tradition}
Let us start with a short review of the common method for the
construction of translation symmetric spin states and the performance
issues connected to it. Consider a quantum spin model with translation
symmetry on a one-dimensional lattice with $n$ sites and periodic
boundary conditions, $\vec{s}_n \equiv \vec{s}_0$,
\begin{equation}\label{eq:ham}
  H = \sum_{i,j} J_{j-i}\ \vec{s}_i\cdot\vec{s}_j\,.
\end{equation}
The Hamiltonian $H$ commutes with the total spin and its components
$S^\alpha=\sum_{i=0}^{n-1} s_i^\alpha$, $\alpha\in\{x,y,z\}$, and with
the translation operator
\begin{equation}\label{eq:trans}
  T: \vec{s}_i \to \vec{s}_{i+1}\,.
\end{equation}
Assuming local spins with amplitude $|\vec{s}_i| = 1/2$, the Hilbert
space of the $n$-site system is the product of $n$ two-dimensional
spaces and has dimension $2^n$. Using the conservation of $S^z$, this
space can be decomposed into $n+1$ subspaces corresponding to the
eigenvalues of $S^z$, $-n/2,-n/2+1,\dots,n/2$. Fixing $u=S^z+n/2$, a
subspace is spanned by all products of $u$ up-spins and $n-u$
down-spins (the two eigenstates of $s_i^z$) and its dimension is
$\binom{n}{u}$, where obviously $\sum_{u=0}^{n} \binom{n}{u} = 2^n$.
On a computer, these states are usually represented as bit patterns of
length $n$, and the above decomposition is equivalent to grouping
patterns according to their digit sum. Understanding bit patterns as
integers defines an order and allows for the construction of ordered
lists, which can be efficiently searched for specific patterns.

The above Hamiltonian also conserves the amplitude of the total spin,
$\vec{S}^2$, but the construction of the corresponding eigenstates is
more involved and rarely adopted in numeric computations on finite
clusters. Instead, lattice symmetries are used to further decompose
spaces of given $S^z$ into smaller subspaces. As the title implies, 
in this work we focus on the translation symmetry.

Since the Hamiltonian $H$ commutes with the translation $T$, the
matrix elements of $H$ between different eigenspaces of $T$
vanish. Thus, if $H$ is expressed in an orthonormal basis of
eigenstates of $T$, the original problem splits into $n$ independent
pieces, each having a dimension roughly a factor of $n$ smaller. The
projection operator
\begin{equation}
  P_k = \frac{1}{n}\sum_{j=0}^{n-1} e^{2\pi i j k/n} T^j
\end{equation}
maps an arbitrary state onto an eigenstate of $T$ with eigenvalue 
$\exp(-2\pi i k/n)$, namely
\begin{equation}
  \begin{aligned}
    T P_k \ket{\psi} & = 
    \frac{1}{n}\sum_{j=0}^{n-1} e^{2\pi i j k/n} T^{j+1} \ket{\psi}\\
    & = \exp(-2\pi i k/n)\,P_k \ket{\psi}\,.
  \end{aligned}
\end{equation}
Here we used $T^n=1$, which also implies $\exp(-2\pi i k) = 1$ and
$k\in\mathbb{Z}$. Due to the periodicity there are only $n$ distinct
projectors and it suffices to consider the momenta
$k=0$, $1$, $\dots$, $(n-1)$.

\begin{figure}
  \begin{center}
    \begin{tikzpicture}[<-,>=stealth,node distance=4mm,semithick,
      dot dashed/.style={dash pattern=on 1pt off 1pt on 3pt off 1pt}]

      \node (n1) {\textbf{0000}};
      \node (n2) [below of=n1] {\textbf{0001}};
      \node (n3) [below of=n2] {0010};
      \node (n4) [below of=n3] {\textbf{0011}};
      \node (n5) [below of=n4] {0100};
      \node (n6) [below of=n5] {\textbf{0101}};
      \node (n7) [below of=n6] {0110};
      \node (n8) [below of=n7] {\textbf{0111}};
      \node (n9) [below of=n8] {1000};
      \node (n10) [below of=n9] {1001};
      \node (n11) [below of=n10] {1010};
      \node (n12) [below of=n11] {1011};
      \node (n13) [below of=n12] {1100};
      \node (n14) [below of=n13] {1101};
      \node (n15) [below of=n14] {1110};
      \node (n16) [below of=n15] {\textbf{1111}};
      
      \path (n1) edge [loop left] (n1);
      
      \path[densely dashed] (n2) edge [bend left=90] (n3);
      \path[densely dashed] (n3) edge [bend right=90] (n5);
      \path[densely dashed] (n5) edge [bend left=90] (n9);
      \path[densely dashed] (n9) edge [bend left=90] (n2);
      
      \path[dot dashed] (n4) edge [bend left=90] (n7);
      \path[dot dashed] (n7) edge [bend right=90] (n13);
      \path[dot dashed] (n10) edge [bend left=90] (n4);
      \path[dot dashed] (n13) edge [bend right=90] (n10);
      
      \path (n6) edge [bend left=90] (n11);
      \path (n11) edge [bend left=90] (n6);
      
      \path[densely dotted] (n8) edge [bend right=90] (n15);
      \path[densely dotted] (n12) edge [bend right=90] (n8);
      \path[densely dotted] (n14) edge [bend left=90] (n12);
      \path[densely dotted] (n15) edge [bend right=90] (n14);
      
      \path (n16) edge [loop left] (n16);
      
    \end{tikzpicture}
  \end{center}
  % \centering\includegraphics{fig1}
  \caption{Decomposition of the set of $S^z$
    eigenstates with $n=4$ into orbits. Arrows indicate the action of
    the translation $T$. The six representatives are shown in bold
    face.}\label{fig:orbits4}
\end{figure}
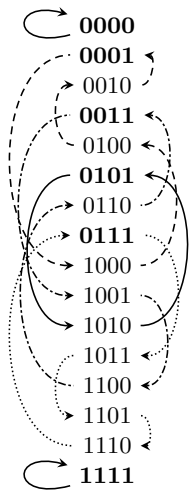

The projector $P_k$ maps states, which are related to each other by an
arbitrary translation, $\ket{\phi} = T^j \ket{\psi}$, onto the
same eigenstate of $T$ (up to a phase factor). To avoid this ambiguity
and to obtain a symmetrised basis we need to partition the set
of all $S^z$ eigenstates, $\mathcal{S}$, into orbits, i.e., disjoint
subsets $\mathcal{S}_r$ that are closed under the translation group,
\begin{equation}
  \forall \ket{\psi} \in \mathcal{S}_r:
  \ T^j \ket{\psi} \in \mathcal{S}_r \,.
\end{equation}
Each orbit can be represented by one of its elements. For instance, if
we use bit patterns to represent $S^z$ eigenstates, we can sort all
elements of an orbit by the corresponding integer values and choose
the smallest as the representative of the orbit. All other members of
this orbit are obtained from the representative by applying all
translations $T^j$, $j=1, \dots, (n-1)$. Figure~\ref{fig:orbits4}
illustrates this decomposition for the case $n=4$, where $\mathcal{S}$
contains $2^4=16$ elements which belong to $6$ disjoint orbits. The
sizes of the orbits differ, since some of the bit patterns are
invariant under non-trivial subgroups of the full translation group of
the $n$-site lattice. The orbit size is then given by $n$ divided by
the order of the subgroup.

Let us denote the set of all representatives with $\mathcal{R}$ and
the size of the orbit of $\ket{r}\in\mathcal{R}$ with
$\omega_r$. Then, for given momentum $k$ a translation symmetric basis
of the Hilbert space is formed by the states
\begin{equation}\label{eq:krstate}
  \ket{k,r} = \sqrt{\nu_{k,r}}\ P_k \ket{r},\ \ket{r}\in\mathcal{R}\,.
\end{equation}
The prefactor $\sqrt{\nu_{k,r}}$ ensures that $\ket{k,r}$ is
normalised, $\langle k,r|k,r\rangle = 1$, or it is zero, which means
that for momentum $k$ the representative $\ket{r}$ does not
contribute to the basis. More precisely,
\begin{equation}\label{eq:krnorm}
  \nu_{k,r} = \begin{cases} 
    0 & \text{if $n/\omega_r$ divides $k$,}\\
    \omega_{r} & \text{otherwise.}
  \end{cases}
\end{equation}
Of course, we can restrict the set $\mathcal{R}$ to states with a
given eigenvalue of $S^z$ (i.e., patterns with given digit sum), which
yields a basis that makes use of both symmetries of $H$, $S^z$, and
$T$.

\section{The Problem}\label{sec:problem}
The recipe for the construction of the symmetrised basis
$\{\ket{k,r}\}$ does not look particularly complicated. However,
following it becomes very time-consuming for large $n$. To find all
representatives $\mathcal{R}$ we can loop over the $2^n$ eigenstates
of $S^z$ in $\mathcal{S}$ (or at least the $\binom{n}{S^z+n/2}$
eigenstates of fixed $S^z$), apply all $n$ translations, and check if
the considered bit pattern has the minimal integer value within its
orbit. If this is the case, it qualifies for the set
$\mathcal{R}$. The process can be improved by memorising bit patterns
that were already encountered in previous orbits and apply the
translations only to new ones. Still, we need to perform of the order
of $2^n$ translations and store all the patterns.

The performance issues become more serious once we start to calculate
matrix elements of the Hamiltonian $H$ with respect to the basis
$\{\ket{k,r}\}$,
\begin{equation}
  \begin{aligned}
    \bra{k,r'} H \ket{k,r}  & = 
    \sqrt{\nu_{k,r'}\nu_{k,r}}\ \bra{r'} P_k H P_k \ket{r} \\
    & = \sqrt{\nu_{k,r'}\nu_{k,r}}\ \bra{r'} P_k H \ket{r}\,.
  \end{aligned}
\end{equation}
The application of $H$ on a representative $\ket{r}$ yields many
different bit patterns---usually their number is some multiple of the
lattice size $n$. In general, these bit patterns are not
representatives, and we need to find the orbit they belong to as well
as the translation, which maps the pattern to the representative of
its orbit. The latter tells us which part of the projector $P_k$
contributes to the matrix element, in particular, which phase
factor. If we apply all translations to all bit patterns generated by
$H$ and then look up the observed representatives in a list, the
construction of the (sparse) matrix representation of $H$ requires
huge amounts of bit operations and processing time. 

In contrast, for lattice models such as the Hubbard model or
electron-phonon models, the Hilbert space is the product of subspaces
which can be symmetrised individually. The subspaces are small enough
such that orbit representatives can be identified through simple table
look-ups, and it is common practise to construct the Hamiltonian
matrix on-the-fly in each step of an iterative calculation. Methods
such as the Lanczos eigenvalue solver~\cite{La50} then need memory
only for a few vectors with the dimension of the Hilbert space, and
huge problems can be studied.

\begin{table}
  \begin{tabular}{rr}
    $n$ & Dimension \\
    \hline
    36 & 252\thsep088\thsep496\\
    38 & 930\thsep138\thsep522\\
    40 &  3\thsep446\thsep167\thsep860\\
    42 & 12\thsep815\thsep663\thsep844\\
    44 & 47\thsep820\thsep447\thsep028\\
    46 & 178\thsep987\thsep624\thsep514
  \end{tabular}
  \vspace{3pt}
  \caption{Problem dimensions for $S^z=0$, $k=0$ as a function of 
    the chain length $n$.}\label{tab:dims}
\end{table}

With the standard approach for quantum spin models this is
impractical. Instead, the matrix has to be kept in memory or stored on
disk. The former limits the accessible system sizes, and the latter is
not efficient either, since disk access is slow and the matrix
dimensions are huge (cf. Table~\ref{tab:dims}). 

The program SPINPACK~\cite{spinpack}, which employs many symmetries and is
frequently used to calculate the lowest eigenstates of spin models,
follows the above strategies, and the problem of identifying the orbit
and representative for a given bit pattern seriously limits its
performance. The authors of the code even considered implementing the
required bit operations with specialised hardware based on
field-programmable gate arrays (FPGA)~\cite{spinpack-fpga}. The
authors of Ref.~\cite{LSS11}, on the other hand, argue that the
overhead for using symmetries outweighs the benefits of the reduced
problem dimensions, and in their large-scale exact diagonalisation
study make no use of translation symmetries.

Below we resolve all these issues and present an answer to the
following problem: \emph{Find a fast and memory-efficient algorithm,
  which for a given arbitrary bit pattern identifies the orbit the
  pattern belongs to and the translation that maps it to the
  representative of this orbit.}

\section{Divide-and-conquer approach}\label{sec:divnconq}

\subsection{The basic idea}
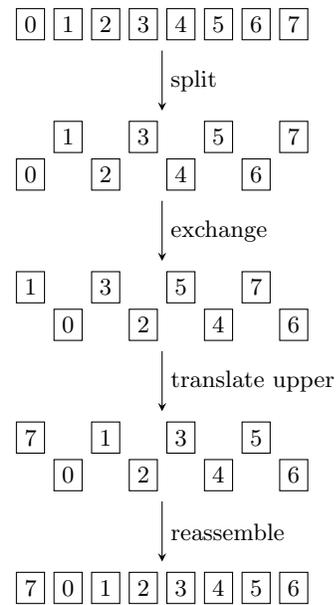
\begin{figure}
  \begin{center}
    \begin{tikzpicture}[scale=0.5,>=stealth]
      % \tikzstyle{site}=[circle,fill=black!25,minimum size=13pt,inner sep=1pt]
      \tikzstyle{site}=[draw,inner sep=3pt]
      
      \foreach \i in {0,...,7} \node[site] at (\i,0) {$\i$};
      
      \draw[->] (3.5,-0.7) -- node[right] {split} (3.5,-2.3);
      
      \foreach \i in {1,3,...,7} \node[site] at (\i,-3) {$\i$};
      \foreach \i in {0,2,...,6} \node[site] at (\i,-4) {$\i$};
      
      \draw[->] (3.5,-4.7) -- node[right] {exchange} (3.5,-6.3);
      
      \foreach \i in {1,3,...,7} \node[site] at (\i-1,-7) {$\i$};
      \foreach \i in {0,2,...,6} \node[site] at (\i+1,-8) {$\i$};
      
      \draw[->] (3.5,-8.7) -- node[right] {translate upper} (3.5,-10.3);
      
      \foreach \i/\v in {1/7,3/1,5/3,7/5} \node[site] at (\i-1,-11) {$\v$};
      \foreach \i in {0,2,...,6} \node[site] at (\i+1,-12) {$\i$};
      
      \draw[->] (3.5,-12.7) -- node[right] {reassemble} (3.5,-14.3);
      
      \foreach \i/\v in {0/7,1/0,2/1,3/2,4/3,5/4,6/5,7/6} \node[site] at (\i,-15) {$\v$};
    \end{tikzpicture}
  \end{center}
  % \centering\includegraphics{fig2}
  \caption{The action of a translation on a lattice decomposed into two intertwined sublattices.}\label{fig:transdeco}
\end{figure}

Let us assume that the number of lattice sites $n$ is even. We can
then divide the set of all sites into two subsets of equal size, such
that the neighbours of a given site all belong to the other sublattice.
The translation $T$ of the entire lattice is then decomposed into two
operations: the exchange of the two sublattices and a translation
within one of the two. In Figure~\ref{fig:transdeco} we illustrate
this concept for a lattice of $n=8$ sites. 

Such a decomposition---termed ``sublattice coding''---was used by
H.\,Q.~Lin~\cite{Li90} to construct the orbit representatives in exact
diagonalisation studies of translation symmetric spin clusters with up
to $32$ sites. Unfortunately, the description of his algorithm is
rather brief and details of the implementation remain
vague. Therefore, the potential of this trick seems to have been
missed (see, e.g., the discussion of Ref.~\cite{Li90} in
Refs.~\cite{Sa10,Lae11}). 

An improved implementation of the Lin decomposition was proposed by
Schulz, Ziman, and Poilblanc in Ref.~\cite{SZP96}. Here an
arbitrary spin state is decomposed into its sublattice states and the
representatives of the sublattice orbits are determined through
look-ups in moderately sized tables.  The operation that maps one
sublattice state to its representative is then applied to the other
sublattice, which in most cases ($80$~\% according to Ref.~\cite{SZP96})
yields the correct orbit representative of the full lattice. However,
an ambiguity remains and additional symmetry operations can be necessary
to identify the correct representative.

In what follows we explain our interpretation and extension of the Lin
approach, which has been used for a couple of years and efficiently
handles very large spin systems. We properly decompose all symmetry
operations and for each given spin state arrive at a unique orbit
representative purely through look-ups in moderately sized
tables. There is no need to apply symmetry operations to spin states
on the full lattice, i.e., the above ambiguity is resolved.

As a starting point, we formalise the above decomposition of lattice
translations by introducing the ``zipper product'' of two bit patterns
$\ket{a}=(a_0,\dots,a_{n/2-1})$ and $\ket{b}=(b_0,\dots,b_{n/2-1})$,
\begin{multline}
  (a_0,\dots,a_{n/2-1})\zp(b_0,\dots,b_{n/2-1}) \\
  := (a_0,b_0,\dots,a_{n/2-1},b_{n/2-1})\,.
\end{multline}
For clarity, we indicate the size of the translated lattice as an
index to the translation operator, $T_n$, and assume that it
translates patterns to the right,
\begin{equation}
  T_n (a_0,\dots,a_{n-1}) = (a_{n-1},a_0,\dots,a_{n-2})\,.
\end{equation}
Then, the above procedure can be summarised as
\begin{equation}\label{eq:transdeco}
  T_n(\ket{a}\zp \ket{b}) = (T_{n/2}\, \ket{b})\zp \ket{a}\,.
\end{equation}
This equation relates the translations on the $n$-site lattice to the
translations on the $n/2$-site sublattice, and we can use it to derive
orbits and representatives of the full lattice from the orbits and
representatives of the sublattice. Multiple application of $T_n$
on a state $\ket{a}\zp\ket{b}$ yields
\begin{equation}
  \begin{aligned}
    T_n^{2j} (\ket{a}\zp \ket{b}) & = (T_{n/2}^j \ket{a})\zp(T_{n/2}^j \ket{b})\,,\\
    T_n^{2j+1} (\ket{a}\zp \ket{b}) & = (T_{n/2}^{j+1} \ket{b})\zp(T_{n/2}^j \ket{a})\,,
  \end{aligned}
\end{equation}
with $j=0,\dots,(n/2-1)$, which illustrates how the orbit of
$\ket{a}\zp \ket{b}$ is built from the sublattice orbits of $\ket{a}$
and $\ket{b}$.

Let us now consider two representatives of the sublattice $\ket{r},
\ket{r'}\in\mathcal{R}_{n/2}$ subject to two conditions: First,
$r<r'$, where the order is defined in terms of the integer value of
the bit patterns. Second, the orbits of both representatives have
maximal size $n/2$, i.e., $T_{n/2}^j\ket{\psi} \ne \ket{\psi}$
$\forall j=1,\dots,(n/2-1)$ and $\ket{\psi}\in\{\ket{r},\ket{r'}\}$. Then,
the $n/2$ states
\begin{equation}\label{eq:genreps}
  \ket{r,r',j} = \ket{r} \zp (T_{n/2}^j \ket{r'})\text{ with }
  j = 0,\dots,(n/2-1)
\end{equation}
are representatives for orbits of the full translation group on the
$n$-site lattice. Since the translation $T_n$ involves the exchange of
the two sublattices (see Eq.~\ref{eq:transdeco}), the orbit of
$\ket{r,r',j}$, which is given by $T_n^i \ket{r,r',j}$ with $i=0,\dots,(n-1)$, contains
all $n^2/2$ states that can be obtained by combining the sublattice
orbits of $\ket{r}$ and $\ket{r'}$, namely
$(T_{n/2}^j\ket{r})\zp(T_{n/2}^l\ket{r'})$ \emph{and}
$(T_{n/2}^l\ket{r'})\zp(T_{n/2}^j\ket{r})$, with
$j,l=0,\dots,(n/2-1)$. This explains the above condition $r<r'$. 

For $r=r'$, the range of $j$ needs to be reduced,
\begin{equation}\label{eq:equreps}
    \ket{r,r,j} = \ket{r} \zp (T_{n/2}^j \ket{r})\text{ with }
    j = 0,\dots,\lfloor (n-1)/4\rfloor\,,
\end{equation}
as otherwise we would count states twice. Note, that for $n/2$ odd one
of the generated representatives is invariant under $T_{n}^{n/2}$,
i.e., the orbit has size $n/2$ only.

The general case, where $r\le r'$ and $\ket{r}$ or $\ket{r'}$ are
invariant under non-trivial subgroups of the $n/2$-site translation
group, leads to further restrictions on the values of $j$. In
addition, the generated representatives $\ket{r,r',j}$ will be
invariant under subgroups of the $n$-site translation group. It is
then convenient to identify all subgroups of the $n/2$-site
translations (which correspond to the divisors of $n/2$) and to
tabulate all possible combinations and the resulting restrictions on
$j$. These tables are small and can also hold other basic details,
like the orbit size $\omega_{r,r',j}$ of $\ket{r,r',j}$, which does
not depend on $r$ and $r'$ directly but only on the maximal subgroups
that $\ket{r}$ and $\ket{r'}$ are invariant under. 

What are the advantages of the decomposition into two sublattices?
First, the number of representatives of the sublattice is
approximately equal to the square root of the number of
representatives of the full lattice, $|\mathcal{R}_{n/2}| \approx
\sqrt{|\mathcal{R}_{n}|}$. Hence, the construction of
$\mathcal{R}_{n}$ from $\mathcal{R}_{n/2}$ is much faster than the
traditional approach we described earlier. Second,
$|\mathcal{R}_{n/2}|$ is small enough, that we can store in memory the
map $\ket{\psi}\to T_{n/2}^j\ket{r}$ from an arbitrary state
$\ket{\psi}\in\mathcal{S}_{n/2}$ to its representative
$\ket{r}\in\mathcal{R}_{n/2}$ and the corresponding exponent
$j$. Moreover, we can use these tables to directly identify the
representative $\ket{r}\in\mathcal{R}_{n}$ and the exponent $j$ for an
arbitrary state $\ket{\psi}\in\mathcal{S}_{n}$ on the full lattice.
Hence, we can solve the problem of Section~\ref{sec:problem} with a
few look-ups in moderately sized tables (typically a few megabytes).

\subsection{Implementation}
We start from a lattice with $n/2$ sites and determine all subgroups
of the translation group generated by $T_{n/2}$. They are given by the
divisors of $n/2$, namely, if $d\mid n/2$ then $T_{n/2}^d$ generates a
subgroup. For example, setting $n/2=4$ we find three subgroups indexed with $g$,
\begin{equation}\label{eq:subgroups4}
  \begin{tabular}{ccc}
    $\quad g\quad$ & $\quad d=\omega_g\quad$ & Example\\
    \hline
    0 & 1 & 0000 \\
    1 & 2 & 0101 \\
    2 & 4 & 0001
  \end{tabular}
\end{equation}
which match the three different orbit types
shown in Figure~\ref{fig:orbits4} and the corresponding orbit sizes
$\omega_g$.

Next, we construct representatives for all orbits in
$\mathcal{S}_{n/2}$. Since we are dealing with only half of the target
lattice size $n$, we can use the approach sketched in the first
paragraph of Section~\ref{sec:problem}. For each representative
$\ket{r}\in\mathcal{R}_{n/2}$, we also determine the maximal subgroup $g$
it is invariant under, i.e., we find the minimal non-zero $d\mid n/2$ such that
$T_{n/2}^d\ket{r} = \ket{r}$. For the example $n/2=4$, we obtain
\begin{equation}\label{eq:reps4}
  \begin{tabular}{cccc}
    $\quad r\quad$ & $\quad\ket{r}\quad$ & $\quad d\quad$ & $\quad g\quad$\\
    \hline
    0 &  0000 & 1 & 0\\
    1 &  0001 & 4 & 2\\
    2 &  0011 & 4 & 2\\
    3 &  0101 & 2 & 1\\
    4 &  0111 & 4 & 2\\
    5 &  1111 & 1 & 0\\
  \end{tabular}
\end{equation}
Having selected a set of representatives $\mathcal{R}_{n/2}$, we can
tabulate the map $\ket{\psi}\to T_{n/2}^j\ket{r}$, i.e., we can
construct an array which takes the integer value of a bit pattern as
the index and returns both, the index $r$ of the corresponding
representative $\ket{r}$ and the exponent $j$. For the example $n/2=4$
this looks as follows:
\begin{equation}\label{eq:reverse4}
  \begin{tabular}{rccc}
    \multicolumn{2}{c}{$\psi\equiv\ket{\psi}$} & $\quad r\quad$ & $\quad j\quad$ \\
    \hline
    0 & 0000 & 0 & 0\\
    1 & 0001 & 1 & 0\\
    2 & 0010 & 1 & 3\\
    3 & 0011 & 2 & 0\\
    4 & 0100 & 1 & 2\\
    5 & 0101 & 3 & 0\\
    6 & 0110 & 2 & 3\\
    7 & 0111 & 4 & 0\\
    8 & 1000 & 1 & 1\\
    9 & 1001 & 2 & 1\\
    10 & 1010 & 3 & 1\\
    11 & 1011 & 4 & 1\\
    12 & 1100 & 2 & 2\\
    13 & 1101 & 4 & 2\\
    14 & 1110 & 4 & 3\\
    15 & 1111 & 5 & 0\\
  \end{tabular}
\end{equation}

Knowing all details about the sublattice with $n/2$ sites, we can
construct a set of tables, which characterise the symmetrised states
of the full $n$-site lattice. Consider an arbitrary state
$\ket{a}\zp\ket{b}$ on the full lattice: For both sublattice states,
$\ket{a}$ and $\ket{b}$, we can immediately look up the indices $r_a$ and
$r_b$ of corresponding representatives $\ket{r_a}$ and $\ket{r_b}$, as
well as the exponents $j_a$ and $j_b$. In addition, given $r_a$ and
$r_b$ we know the subgroups $g_a$ and $g_b$ of the representatives
$\ket{r_a}$ and $\ket{r_b}$. The only information missing for locating
$\ket{a}\zp\ket{b}$ within its orbit,
\begin{equation}\label{eq:fullorbit}
  \ket{a}\zp\ket{b} = T_{n}^i \ket{r,r',j}
   = T_{n}^i \left[\ket{r}\zp (T_{n/2}^j\ket{r'})\right]\,,
\end{equation}
are the exponents $i$ and $j$. These exponents depend only on the
values of $j_a$, $j_b$, $g_a$, and $g_b$, and on the order of $r_a$ and
$r_b$, namely, whether $r_a<r_b$, $r_a=r_b$ or $r_a>r_b$. When we
defined the representatives $\ket{r,r',j}$ of the full lattice in
Eq.~(\ref{eq:genreps}), we demanded $r<r'$. Thus, if we encounter
$r_a>r_b$, then $\ket{a}\zp\ket{b}$ is created from the representative
$\ket{r_b,r_a,j}$ by a translation $T_n^i$ with odd exponent $i$. For
$r_a<r_b$, the representative is $\ket{r_a,r_b,j}$ and the exponent $i$
is even. For $r_a=r_b$ other restrictions apply, as discussed in
the paragraph of Eq.~(\ref{eq:equreps}). We can tabulate all possible
cases in three arrays,
\begin{equation}\label{eq:explookup}
  \begin{aligned}
    e^{<}&: j_a, j_b, g_a, g_b \to i,j\,,\\
    e^{=}&: j_a, j_b, g \to i,j\,,\\
    e^{>}&: j_a, j_b, g_a, g_b \to i,j\,.
  \end{aligned}
\end{equation}
At first glance these four-dimensional arrays appear large, but the
$j$ indices take only $n/2$ different values, and the $g$ indices even
fewer. In the Appendices~\ref{sec:elt} to~\ref{sec:egt} we show the maps
$e^{<}$, $e^{=}$, and $e^{>}$ for the lattice with $n=8$ sites. To
build the arrays we perform a double-loop over the subgroups in
Eq.~(\ref{eq:subgroups4}), which fixes $g_a$ and $g_b$. Then, for each
combination of subgroups we pick two matching representatives, $r_a$
and $r_b$, and loop in reverse order over $i$ and $j$ in
Eq.~(\ref{eq:fullorbit}). Looking up the exponents $j_a$ and $j_b$ of
the resulting state $\ket{a}\zp\ket{b}$ in Eq.~(\ref{eq:reverse4})
completes the data required for the arrays. In particular, for each
input $j_a$, $j_b$, $g_a$, and $g_b$, the stored values of $i,j$ are the
minimal ones.

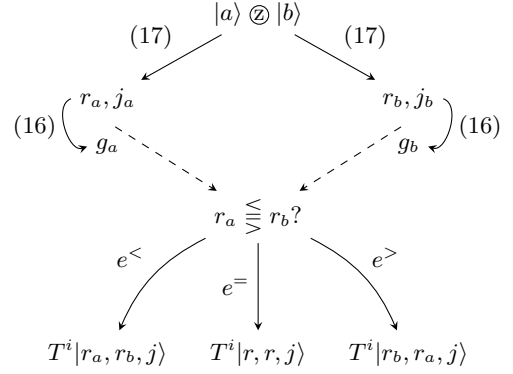
\begin{figure}
  \begin{center}
    \begin{tikzpicture}[>=stealth]
      \node (S) at (0,9.5) {$\ket{a}\zp\ket{b}$};
      \node[anchor=base] (Ra) at (-2,8.3) {$r_a, j_a$};
      \node[anchor=base] (Rb) at ( 2,8.3) {$r_b, j_b$};
      \draw [->] (S) -- node[anchor=south east] {(\ref{eq:reverse4})} (Ra);
      \draw [->] (S) -- node[anchor=south west] {(\ref{eq:reverse4})} (Rb);
      \node[anchor=base] (ga) at (-2,7.7) {$g_a$};
      \node[anchor=base] (gb) at ( 2,7.7) {$g_b$};
      \path (Ra) edge [->,bend right=90] node[anchor=east] {(\ref{eq:reps4})} (ga);
      \path (Rb) edge [->,bend left=90]  node[anchor=west] {(\ref{eq:reps4})} (gb);
      \node (comp) at (0,6.8) {$r_a\lesseqqgtr r_b?$};
      \draw [->,dashed] (-1.9,8) -- (comp);
      \draw [->,dashed] ( 1.9,8) -- (comp);
      \node (Sl) at (-2,5) {$T^i\ket{r_a,r_b,j}$};
      \node (Sq) at ( 0,5) {$T^i\ket{r,r,j}$};
      \node (Sg) at ( 2,5) {$T^i\ket{r_b,r_a,j}$};
      \draw (comp) edge [->,bend right=20] node[anchor=south east] {\hyperref[sec:elt]{$e^{<}$}} (Sl);
      \draw [->] (comp) -- node[anchor=east] {\hyperref[sec:equ]{$e^{=}$}} (Sq);
      \draw (comp) edge [->,bend left=20] node[anchor=south west] {\hyperref[sec:egt]{$e^{>}$}} (Sg);
    \end{tikzpicture}
  \end{center}
  % \centering\includegraphics{fig3}
  \caption{Schematic view of the table look-ups (solid lines) required to 
    locate an arbitrary state $\ket{a}\protect\zp\ket{b}$
    within its orbit.}\label{fig:lookup}
\end{figure}

In Figure~\ref{fig:lookup} we summarise the algorithm to identify
both the orbit of an arbitrary state $\ket{a}\zp\ket{b}$ and its
translation relative to the orbit's representative. Let us remark
that, in general, the representative $\ket{r,r',j}$ is not the
state with minimal integer value within its orbit. Using direct
table look-ups, this property is no longer needed, and in our programs
the components of zipped states are usually stored in separate
variables.

The tables $e^{<}$ and $e^{=}$ can also be used to identify the values
of $j$, for which $\ket{r,r',j}$ is a valid representative. In this
case the set $(j_a=0, j_b=j, g_a, g_b)$ is mapped to $(i=0, j)$, i.e.,
$\ket{r,r',j}$ is not part of the orbit of some other representative
$\ket{r,r',j'}$ with $j'<j$. We can store this information together
with the size of the orbit of $\ket{r,r',j}$, which depends only on
the corresponding subgroups $g$ and $g'$. Similar to the previous
arrays, we need to distinguish two cases, $\omega_{g,g',j}^{<}$ for
$r<r'$ and $\omega_{g,j}^{=}$ for $r=r'$. If $j$ is invalid, we set
$\omega$ to zero, otherwise it will have some integer value $d\mid
n$. In Appendix~\ref{sec:omegas} we list the latter quantities for $n=8$.

In analogy to Eqs.~(\ref{eq:krstate}) and~(\ref{eq:krnorm}), we now
know the normalised, translation-symmetric basis states of the
$n$-site lattice,
\begin{equation}\label{eq:fullkrstate}
  \ket{k,r,r',j} =\sqrt{\nu_{k,r,r',j}}\, P_k \ket{r,r',j}\,,
\end{equation}
where
\begin{equation}\label{eq:fullkrnorm}
  \nu_{k,r,r',j} = \begin{cases} 
    0 & \text{if $n/\omega$ divides $k$}\\
    \omega & \text{otherwise}
  \end{cases}
\end{equation}
and $\omega$ is $\omega_{g,g',j}^{<}$ or $\omega_{g,j}^{=}$,
respectively. Note that for large $n$ the number of representatives
with $\omega<n$ is negligible compared to those with $\omega=n$
introduced in Eq.~(\ref{eq:genreps}). Therefore, in a practical
calculation a loop over the whole basis can include all $r\le r'$ and
$j=0,\dots,(n/2-1)$, and the few inactive states with $\omega=0$ will
waste hardly any resources.

In the preceding paragraphs we did not take into account the $S^z$
symmetry of the original spin model~(\ref{eq:ham}). However, its
inclusion is easy: When constructing the representatives of the
sublattice, $\mathcal{R}_{n/2}$, we also calculate the $S^z$
eigenvalue of each $\ket{r}$. Then, for the representatives of the
full lattice, $\ket{r,r',j}$, we combine only those $r$ and $r'$,
whose spin values add to the desired $S^z$ of the full lattice. This
requires a little extra book keeping, but does not affect the overall
performance.

\section{Generalisations}

\subsection{Two-dimensional lattices}
\begin{figure}
  \begin{center}
    \begin{tikzpicture}[scale=0.7]
      \foreach \i in {0,2} \fill[black!20!white] (0,\i) rectangle (5, \i+1);
      \foreach \n in {0,...,19} {
        \pgfmathsetmacro{\i}{mod(\n,5)};
        \pgfmathsetmacro{\j}{floor(\n/5)};
        \node at (\i+0.5,\j+0.5) {$\n$};
        \draw (\i,\j) rectangle (\i+1, \j+1);
      }
    \end{tikzpicture}
    \hspace{7mm}
    \begin{tikzpicture}[scale=0.7]
      \draw[thick] (0.5, 0.5) -- (3.5, 1.5) -- (2.5, 4.5) -- (-0.5, 3.5) -- cycle;
      \path[fill=black!20!white, draw=black] (0, 0) rectangle (1, 1);
      \node at (0.5,0.5) {$0$};
      \foreach \n in {1,...,9} {
        \pgfmathsetmacro{\i}{floor((\n-1)/3)};
        \pgfmathsetmacro{\j}{mod(\n-1,3)+1};
        \pgfmathsetmacro{\c}{20*mod(\n-1,2)};
        \path[fill=black!\c!white, draw=black] (\i,\j) rectangle (\i+1, \j+1);
        \node at (\i+0.5,\j+0.5) {$\n$};
      }
    \end{tikzpicture}    
  \end{center}
  % \centering\includegraphics{fig4}
  \caption{The decompositions of lattices with $20 = 5\times 4$ and
    with $10 = 3^2 + 1^2$ sites into two
    sublattices.}\label{fig:2Ddecomp}
\end{figure}
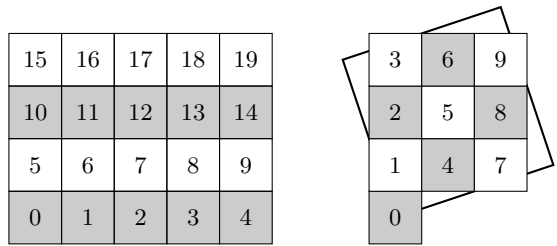

Up to now we considered only one-dimensional lattices, but the
generalisation to two dimensions is straightforward. Again
we demand $n=n_x\times n_y$ to be even. Hence, one or both of $n_x$
and $n_y$ are even, and we can apply the decomposition into
sublattices along one of the two space directions. Another option is
the decomposition into a chequerboard pattern, which can also be used
for quadratic clusters with rotated unit cell~\cite{OB78} and an even
number of sites fulfilling $n=n_1^2 + n_2^2$ with
$n_1,n_2\in\mathbb{Z}$. The main condition for the decomposition is
that the sublattices each have the same translation group. In
Figure~\ref{fig:2Ddecomp} we show the lattice with $20=5\times 4$
sites decomposed along the $y$-direction and the lattice with
$10=3^2+1^2$ sites decomposed in chequerboard fashion.

The construction of representatives for the orbits of the full-lattice
translation group then follows the route described in
Section~\ref{sec:divnconq}. Merely the number and structure of the
subgroups of the translation group differs slightly, since now the
group is generated by two commuting elementary translations $T_x$ and
$T_y$. Also, the condition for vanishing norm $\nu_{k,r,r',j}$ is more
complicated and will usually be tabulated.

\subsection{Reflection symmetry}
Apart from being translation symmetric, most of the considered quantum
spin models are also invariant under reflections, i.e., the full
lattice symmetry is described by the dihedral group or, in higher
dimensions, by products thereof. In one dimension the reflection
operator reads
\begin{equation}
  R: \vec{s}_i \to \vec{s}_{n-1-i}\,.
\end{equation}
It is fully compatible with the lattice decomposition introduced in
Section~\ref{sec:divnconq}, since $R$ can be written as reflections
of both sublattices and exchange of the two,
\begin{equation}
  R(\ket{a}\zp\ket{b}) = (R\ket{b})\zp(R\ket{a})\,.
\end{equation}
Hence, the reflections can be incorporated into the divide-and-conquer
approach and used for a further reduction of the Hilbert space
dimension, or to make the matrix representation real~\cite{Sa10}.

\begin{table*}
  \begin{tabular}{ccc|r|r|crr|r}
    $n$ & $S_z$ & $\dfrac{2\pi k}{n}$ & \multicolumn{1}{c|}{Dimension} & \multicolumn{1}{c|}{$E_{\text{min}}$} 
    & Cores & \multicolumn{1}{c}{Memory} 
    & \multicolumn{1}{c|}{$\dfrac{\text{Time}}{\text{MVM}}$}
    & \multicolumn{1}{c}{$\genfrac{}{}{0pt}{0}{\text{SPINPACK}}{\text{build matrix}}$}\\[2mm]
    \hline
    $1\times 32$ & 0  & $0$ & 18\thsep784\thsep170 & -14.2065274389 & 1 & 290~MB &  10.76~s & $\approx 280$~s\\
    $1\times 34$ & 0  & $\pi$ & 68\thsep635\thsep478 & -15.0912964656 & 1 & 1~GB &  50.03~s & $\approx 1100$~s\\ 
    $1\times 36$ & 0  & $0$ & 252\thsep088\thsep496 & -15.9762358220 & 1 & 3.8~GB &  225.15~s & $\approx 4600$~s\\
    $6\times 6$ & 0  & $(0,0)$ & 252\thsep091\thsep362 & -24.4393973993 & 1 & 3.8~GB &  342.43~s\\
    \hline 
    $1\times 36$ & 0 & $0$ &  252\thsep088\thsep496 & -15.9762358220 & 16 & 7.18~GB &  65.32~s\\
    $6\times 6$ & 0 & $(0,0)$ & 252\thsep091\thsep362 & -24.4393973993 & 16 & 7.20~GB & 112.64~s\\
    $1\times 38$ & 0 & $\pi$ & 930\thsep138\thsep522 & -16.8613184638 & 16 & 26.2~GB &  270.63~s\\
    $1\times 40$ & 0 & $0$ & 3\thsep446\thsep167\thsep860 & -17.7465227882 & 24 & 97.1~GB & \ 1299.15~s\ \\
    \hline
    $1\times 40$ & 0 & $0$ & 3\thsep446\thsep167\thsep860 & -17.7465227883 & 64 & 76.1~GB & 174.53~s\\
    $1\times 42$ & 0 & $\pi$ & 12\thsep815\thsep663\thsep844 & -18.6318313306 & 256 & 249~GB & 207.19~s\\
    $1\times 44$ & 0 & $0$ & 47\thsep820\thsep447\thsep028 & -19.5172298175 & 1024 & 1209~GB & 322.60~s\\
    $1\times 46$ & 0 & $\pi$ &\ 178\thsep987\thsep624\thsep514\ &\ -20.4027064699\ & 1024 & 3522~GB & 1172.84~s\\
  \end{tabular}
  \vspace{3pt}
  \caption{Lanczos calculations of the ground state of the Heisenberg model using 
    one core of a desktop computer with Intel Xeon 5150 processors at 2.66~GHz, 
    many cores of a compute server with 8 quad-core Opteron 8384 processors at 2.7~GHz,
    and several nodes of a high-performance cluster with Power6 processors at 4.7~GHz.
    The eighth column shows the time required for one matrix vector multiplication (MVM), which 
    includes on-the-fly generation of the Hamiltonian matrix. For comparison, in the last column
    we list the time SPINPACK 2.43 needs to generate the matrix (stored in RAM for these system sizes).
    }\label{tab:lanczostiming}
\end{table*}

\subsection{Odd lattice size}
The key prerequisite for the decompositions presented in the preceding
sections, is the even number of lattice sites. In practise, this
condition is not particularly restrictive, since many of the quantum
spin models studied are anti-ferromagnetic. Fitting long-range order
or correlations of this type into a finite cluster usually requires
even $n$. However, lattices with an odd number of sites could be of
interest for certain interaction types, geometries, or spin amplitudes
other than one-half.

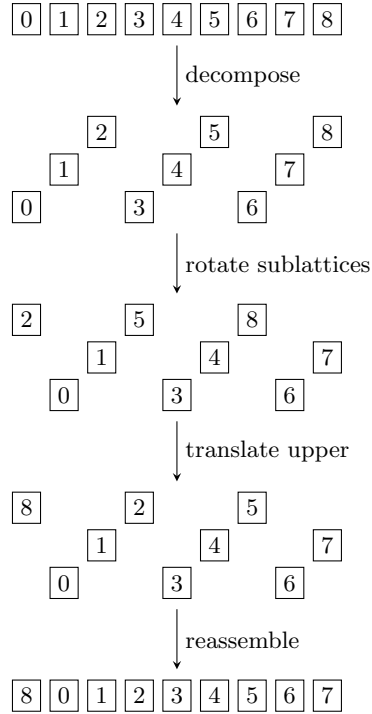
\begin{figure}
  \begin{center}
    \begin{tikzpicture}[scale=0.5,>=stealth]
      \tikzstyle{site}=[draw,inner sep=3pt]
      
      \foreach \i in {0,...,8} \node[site] at (\i,0) {$\i$};
      
      \draw[->] (4.0,-0.7) -- node[right] {decompose} (4.0,-2.3);
      
      \foreach \i in {2,5,...,8} \node[site] at (\i,-3) {$\i$};
      \foreach \i in {1,4,...,7} \node[site] at (\i,-4) {$\i$};
      \foreach \i in {0,3,...,6} \node[site] at (\i,-5) {$\i$};
      
      \draw[->] (4.0,-5.7) -- node[right] {rotate sublattices} (4.0,-7.3);
      
      \foreach \i in {2,5,...,8} \node[site] at (\i-2,-8) {$\i$};
      \foreach \i in {1,4,...,7} \node[site] at (\i+1,-9) {$\i$};
      \foreach \i in {0,3,...,6} \node[site] at (\i+1,-10) {$\i$};
      
      \draw[->] (4.0,-10.7) -- node[right] {translate upper} (4.0,-12.3);
      
      \foreach \i/\v in {2/8,5/2,8/5} \node[site] at (\i-2,-13) {$\v$};
      \foreach \i in {1,4,...,7} \node[site] at (\i+1,-14) {$\i$};
      \foreach \i in {0,3,...,6} \node[site] at (\i+1,-15) {$\i$};

      \draw[->] (4.0,-15.7) -- node[right] {reassemble} (4.0,-17.3);
      
      \foreach \i/\v in {0/8,1/0,2/1,3/2,4/3,5/4,6/5,7/6,8/7} \node[site] at (\i,-18) {$\v$};
    \end{tikzpicture}
  \end{center}
  % \centering\includegraphics{fig5}
  \caption{The action of a translation on a lattice decomposed into three 
    sublattices of equal size.}\label{fig:deco3}
\end{figure}

As long as $n$ is not prime, we can take a small factor $m\mid n$ and
split the lattice into $m$ equal sublattices, such that the left and
the right neighbor of a site belongs to the previous and the next
sublattice, respectively. An arbitrary translation of the full lattice
then corresponds to a cyclic permutation of the sublattices and
internal translations within the sublattices.

Consider, for instance, a lattice where the number of sites is a
multiple of $3$. We can decompose this lattice into 3 sublattices, as
illustrated in Figure~\ref{fig:deco3}. Now, the translation of the full
lattice by a single site is equivalent to a cyclic permutation of the
three sublattices and an internal translation within one of the three,
or
\begin{equation}
  T_n(\ket{a}\zp\ket{b}\zp\ket{c}) = (T_{n/3}\ket{c})\zp\ket{a}\zp\ket{b}\,.
\end{equation}
Knowing the representatives $\ket{r}$ for the sublattice with $n/3$ sites,
we can build representatives for the full lattice,
\begin{equation}
  \ket{r,r',r'',j',j''} = \ket{r}\zp(T_{n/3}^{j'}\ket{r'})\zp(T_{n/3}^{j''}\ket{r''})\,.
\end{equation}
The restrictions needed to avoid double counting are more intricate,
compared to the bi-partition. There are 6 permutations of 3 objects,
and the cycle $(123)$ connects the even and the odd permutations among
each other. Thus, starting from $r\le r' \le r''$ or $r>r'>r''$ we can
reach all possible combinations of sublattice states. The exponents,
in general, can take all values $j',j''=0,\dots,(n/3-1)$, but some
will be switched off with appropriate norm factors, if two or all
representatives are equal or have a higher symmetry. When we construct
the corresponding tables of the orbit sizes
$\omega_{g,g',g'',j',j''}$, we need to differentiate between a number
of different cases. The three states $r$, $r'$, and $r''$ can all be
different and arranged in ascending or descending order, there can be
equal pairs, or all three can be the same. Similarly, the three tables
$e^{<}$, $e^{=}$, and $e^{>}$, which for the bi-partite lattice were
sufficient to identify the orbit and the phase factor of an arbitrary
state on the full lattice, now generalise to a whole set of tables
covering all possible orderings of $r$, $r'$, and $r''$.

As yet we did not have a good incentive to study lattices with an odd
number of sites and, therefore, cannot comment on the performance of
this setup. An implementation of the decomposition into three
sublattices appears feasible, but the benefits of decompositions into
five or more sublattices seem to be rather limited.

\subsection{Higher spin}
The translation symmetry of the lattice and the structure of the local
Hilbert space at each site are more or less independent. Therefore,
the construction of the translation symmetric basis can easily be
extended to systems with spins of amplitude larger than $1/2$. The
efficient storage of the map from sublattice states to sublattice
representatives, $\ket{\psi}\to T_{n/2}^j\ket{r}$, may require some
care. Otherwise, all steps of the algorithm work as described above.

\section{Performance}

We implemented the divide-and-conquer approach for spin-$1/2$
chains and rectangular two-dimensional lattices ($n=n_x\times n_y$)
already a few years ago, and used it mainly for the study of
correlation functions. The latter can be efficiently calculated using
Chebyshev expansion methods~\cite{WWAF06,WF08b}, which at their
core require fast matrix vector multiplications. For example, we
calculated a set of static correlation functions~\cite{BDGKSW08} and
the dynamic ESR-response~\cite{BGKKW11,BGKKW12} of the one-dimensional
XXZ model at finite temperature and finite magnetic field.
  
Of course, the described basis construction can also be used in
Lanczos calculations of low-energy eigenstates. To give an impression
of the performance of the algorithm, in Table~\ref{tab:lanczostiming}
we show the time and memory consumption of several ground-state
calculations for the Heisenberg model on one- and two-dimensional
lattices. Taking into account the translation and the $S_z$
symmetries, the Hamilton matrix is computed on-the-fly in each
iteration. For the momenta considered the matrix is real. Systems with
up to $36$ sites can be simulated on desktop computers or powerful
laptops, as illustrated by the single-core data for an older Xeon
CPU. For systems with up to $40$ sites, we use a compute server with
eight quad-core CPUs, and on a decent high-performance
cluster~\cite{top500} we are able to handle systems with $46$ sites,
corresponding to a matrix dimension of $1.8\times 10^{11}$. The main
limiting parameter for these calculations is the memory required for
two double vectors with the dimension of the Hilbert space. On the
largest clusters currently available one could certainly study systems
with $50$ sites, which requires approximately $40$~TB of memory and
is well below present limits. 

A direct comparison of our timings with SPINPACK is difficult, since
this code usually precomputes the entire Hamilton matrix and stores it
in memory or on disk for later use in the Lanczos recursion. In the
last column of Table~\ref{tab:lanczostiming} we show matrix generation
times for not too large systems, where the matrix fits into available
memory. These calculations take much longer than the matrix vector
multiplication (MVM) in our approach, which includes on-the-fly matrix
generation.

\section{Summary}

We present an efficient algorithm to construct translation symmetric
basis states for quantum spin models on finite lattices with periodic
boundary conditions. The approach extends an old trick by
H.\,Q.~Lin~\cite{Li90} and employs a divide-and-conquer strategy, such
that direct table look-ups can be used to map an arbitrary spin state
to its orbit with respect to the translation group. The Hamiltonian
matrix, which in iterative calculations like Lanczos or Chebyshev
expansion needs to be applied repeatedly to a few quantum states, can
then be constructed on-the-fly. This saves large amounts of memory or
disk space and considerably increases the system size accessible to
these types of simulations.

We thank Rechenzentrum Garching of the Max Planck Society for providing
computing time on their high-performance clusters.

While this article was under review we learned of unpublished 
exact diagonalisation results~\cite{laeuchli12} for systems with $48$ spins 
(dimension $2.5\times 10^{11}$), which were obtained with an extended version 
of the method in Ref.~\cite{SZP96}.

% \clearpage

\bibliography{ref,special}
\bibliographystyle{apsrev4-1}

\appendix
\begin{widetext}
\section{Tables for $n=8$}

\subsection{The map $e^{<}: j_a, j_b, g_a, g_b \to (i,j)$}\label{sec:elt}
\begin{center}
%  {\small
  \begin{tabular}{cc|C{7mm}C{7mm}C{7mm}C{7mm}|C{7mm}C{7mm}C{7mm}C{7mm}|C{7mm}C{7mm}C{7mm}C{7mm}}
    $e^{<}$ & $g_b$ & \multicolumn{4}{c|}{0} & \multicolumn{4}{c|}{1} & \multicolumn{4}{c}{2}\\
    $g_a$ & $j_a \backslash j_b$ & $ 0 $ & $ 1 $ & $ 2 $ & $ 3 $ & $ 0 $ & $ 1 $ & $ 2 $ & $ 3 $ & $ 0 $ & $ 1 $ & $ 2 $ & $ 3 $\\
    \hline
    \multirow{4}{*}{0} & 0 & (0,0) & - & - & - & (0,0) & (2,0) & - & - & (0,0) & (2,0) & (4,0) & (6,0)\\
     & 1 & - & - & - & - & - & - & - & - & - & - & - & -\\
     & 2 & - & - & - & - & - & - & - & - & - & - & - & -\\
     & 3 & - & - & - & - & - & - & - & - & - & - & - & -\\
    \hline
    \multirow{4}{*}{1} & 0 & (0,0) & - & - & - & (0,0) & (0,1) & - & - & (0,0) & (0,1) & (4,0) & (4,1)\\
     & 1 & (2,0) & - & - & - & (2,1) & (2,0) & - & - & (6,1) & (2,0) & (2,1) & (6,0)\\
     & 2 & - & - & - & - & - & - & - & - & - & - & - & -\\
     & 3 & - & - & - & - & - & - & - & - & - & - & - & -\\
    \hline
    \multirow{4}{*}{2} & 0 & (0,0) & - & - & - & (0,0) & (0,1) & - & - & (0,0) & (0,1) & (0,2) & (0,3)\\
     & 1 & (2,0) & - & - & - & (2,1) & (2,0) & - & - & (2,3) & (2,0) & (2,1) & (2,2)\\
     & 2 & (4,0) & - & - & - & (4,0) & (4,1) & - & - & (4,2) & (4,3) & (4,0) & (4,1)\\
     & 3 & (6,0) & - & - & - & (6,1) & (6,0) & - & - & (6,1) & (6,2) & (6,3) & (6,0)\\
  \end{tabular} %}
\end{center}

\subsection{The map $e^{=}: j_a, j_b, g \to (i,j)$}\label{sec:equ}
\begin{center}
  \begin{tabular}{c|C{7mm}C{7mm}C{7mm}C{7mm}|C{7mm}C{7mm}C{7mm}C{7mm}|C{7mm}C{7mm}C{7mm}C{7mm}}
    & \multicolumn{4}{c|}{$g=0$} & \multicolumn{4}{c|}{$g=1$} & \multicolumn{4}{c}{$g=2$}\\
    $j_a\backslash j_b$ & $ 0 $ & $ 1 $ & $ 2 $ & $ 3 $ & $ 0 $ & $ 1 $ & $ 2 $ & $ 3 $ & $ 0 $ & $ 1 $ & $ 2 $ & $ 3 $\\
    \hline
    0 & (0,0) & - & - & -& (0,0) & (3,0) & - & -& (0,0) & (0,1) & (5,1) & (7,0)\\
    1 & - & - & - & -& (1,0) & (2,0) & - & -& (1,0) & (2,0) & (2,1) & (7,1)\\
    2 & - & - & - & -& - & - & - & -& (1,1) & (3,0) & (4,0) & (4,1)\\
    3 & - & - & - & -& - & - & - & -& (6,1) & (3,1) & (5,0) & (6,0)\\
  \end{tabular}
\end{center}

\subsection{The map $e^{>}: j_a, j_b, g_a, g_b \to (i,j)$}\label{sec:egt}
\begin{center}
  % {\small
  \begin{tabular}{cc|C{7mm}C{7mm}C{7mm}C{7mm}|C{7mm}C{7mm}C{7mm}C{7mm}|C{7mm}C{7mm}C{7mm}C{7mm}}
    $e^{>}$ & $g_a$ & \multicolumn{4}{c|}{0} & \multicolumn{4}{c|}{1} & \multicolumn{4}{c}{2}\\
    $g_b$ & $j_b \backslash j_a$ & $ 0 $ & $ 1 $ & $ 2 $ & $ 3 $ & $ 0 $ & $ 1 $ & $ 2 $ & $ 3 $ & $ 0 $ & $ 1 $ & $ 2 $ & $ 3 $\\
    \hline
    \multirow{4}{*}{0} & 0 & (1,0) & - & - & - & (3,0) & (1,0) & - & - & (7,0) & (1,0) & (3,0) & (5,0)\\
    & 1 & - & - & - & - & - & - & - & - & - & - & - & - \\
    & 2 & - & - & - & - & - & - & - & - & - & - & - & - \\
    & 3 & - & - & - & - & - & - & - & - & - & - & - & - \\
    \hline
    \multirow{4}{*}{1} & 0 & (1,0) & - & - & - & (1,1) & (1,0) & - & - & (5,1) & (1,0) & (1,1) & (5,0)\\
    & 1 & (3,0) & - & - & - & (3,0) & (3,1) & - & - & (7,0) & (7,1) & (3,0) & (3,1)\\
    & 2 & - & - & - & - & - & - & - & - & - & - & - & - \\
    & 3 & - & - & - & - & - & - & - & - & - & - & - & - \\
    \hline
    \multirow{4}{*}{2} & 0 & (1,0) & - & - & - & (1,1) & (1,0) & - & - & (1,3) & (1,0) & (1,1) & (1,2)\\
    & 1 & (3,0) & - & - & - & (3,0) & (3,1) & - & - & (3,2) & (3,3) & (3,0) & (3,1)\\
    & 2 & (5,0) & - & - & - & (5,1) & (5,0) & - & - & (5,1) & (5,2) & (5,3) & (5,0)\\
    & 3 & (7,0) & - & - & - & (7,0) & (7,1) & - & - & (7,0) & (7,1) & (7,2) & (7,3)\\
  \end{tabular} %}
\end{center}

\subsection{Sizes of the orbits represented by $\ket{r,r',j}$}\label{sec:omegas}
\begin{center}
  \begin{tabular}{cc|ccc|ccc|ccc|ccc}
    & & \multicolumn{9}{c|}{$\omega_{g,g',j}^{<}$} & \multicolumn{3}{c}{$\omega_{g,j}^{=}$}\\
    \hline
    & $g$ & \multicolumn{3}{c|}{0} & \multicolumn{3}{c|}{1} & \multicolumn{3}{c|}{2} &  $\  0 \ $ & $\  1 \ $ & $\  2 \ $ \\
    $\ j$ & $g'$ & $\  0 \ $ & $\  1 \ $ & $\  2 \ $ &  $\  0 \ $ & $\  1 \ $ & $\  2 \ $ & $\  0 \ $ & $\  1 \ $ & $\  2 \ $ &  $\ 0\ $ & $\ 1\ $ & $\ 2\ $ \\
    \hline
    0 & & 2 & 4 & 8 & 4 & 4 & 8 & 8 & 8 & 8 & 1 & 4 & 8\\
    1 & & 0 & 0 & 0 & 0 & 4 & 8 & 0 & 8 & 8 & 0 & 0 & 8\\
    2 & & 0 & 0 & 0 & 0 & 0 & 0 & 0 & 0 & 8 & 0 & 0 & 0\\
    3 & & 0 & 0 & 0 & 0 & 0 & 0 & 0 & 0 & 8 & 0 & 0 & 0\\
  \end{tabular}
\end{center}

\end{widetext}

\end{document}